# Coherent manipulation of nitrogen vacancy centers in 4H silicon carbide with resonant excitation

Zhao Mu[1], S.A.Zargaleh[1], H. J. von Bardeleben[2], Johannes E. Froch[3], Hongbing Cai[1], Xinge Yang[1], Jianqun Yang[4], Xingji Li[4], Igor Aharonovich[3], Weibo Gao[1,5]

[1]*Division of Physics and Applied Physics, School of Physical and Mathematical Sciences, Nanyang Technological University, 637371, Singapore*
[2]*Sorbonne Université, Campus Pierre et Marie Curie, Institut des Nanosciences de Paris, 4, place Jussieu, 75005 Paris, France*
[3]*School of Mathematical and Physical Sciences, University of Technology Sydney, Ultimo, New South Wales 2007, Australia*
[4]*School of Materials Science and Engineering, Harbin institute of Technology, Harbin 15000, P.R.China*
[5]*The Photonics Institute and Centre for Disruptive Photonic Technologies, Nanyang Technological University, 637371, Singapore*

Silicon carbide (SiC) has become a key player in realization of scalable quantum technologies due to its ability to host optically addressable spin qubits and wafer-size samples. Here, we have demonstrated optically detected magnetic resonance (ODMR) with resonant excitation, which clearly measured the ground state energy levels of the NV centers in 4H-SiC. Coherent manipulation of NV centers in SiC has been achieved with Rabi and Ramsey oscillations. Finally, we show the successful generation and characterization of single nitrogen vacancy (NV) center in SiC employing ion implantation. Our results are highlighting the key role of NV centers in SiC as a potential candidate for quantum information processing.

Color centers in wide bandgap materials have been studied intensively over the past few decades as potential building blocks for quantum information processing applications [1-4]. Amongst a variety of candidates, the NV center in diamond has garnered most attention with applications in quantum sensing [5,6] and quantum memories [1,7]. However, the NV optical properties are not ideal for applications at the telecom range, where industry communications are established. Furthermore, the diamond as a host is an expensive material that only comes in small wafer sizes.

Silicon carbide, on the other hand, is a mature industrial semiconductor material [2,3,8,9]. Similarly to diamond, it can host isolated defects that can be harnessed as spin qubits in both the visible and the infrared spectral range [8,10-27]. In addition, the coherent manipulation of the electron spin of several defects, including the silicon vacancy [19,21,28-31] and divacancy [8,10,11,31-34] have been demonstrated, thus making defects in SiC appealing for quantum applications.

Recently, another family of defects, namely the NV centers in SiC, has attracted a considerable attention, due to its promising magnetic and optical properties. NV center in 4H-SiC has one silicon vacancy with one nitrogen atom as the nearest neighbor ($V_{Si}N_C$). In the charged state, it has the same electronic structure as NV centers in diamond with an $^3A_2$ ground state and an $^3E$ excited state, sharp zero-phonon PL lines and an intermediate singlet state allowing optically induced ground-state spin polarization. These defects are particularly interesting because their emission is closer to the telecom wavelength band and thus offer unique perspectives for quantum communications.

In this work, we perform ODMR spectroscopy with high resolution resonant excitation of each of the four different types of NV centers in SiC. Taking advantage of resonant excitation, the ground state splitting of the NV centers has been clearly measured at both zero and non-zero magnetic fields. Moreover, Rabi Oscillation and Ramsey fringes for NV center ensembles have been demonstrated, revealing a spin dephasing time of $T_2$*=0.4μs. Finally, we have shown that single NV center can be generated with low-dose ion implantation.

Due to the lower crystal symmetry of 4H-SiC as compared to diamond, four different NV centers coexist with similar but distinct optical and magnetic properties, as shown in Figure 1(a). Here we label different types of NV centers as kh, hh, kk and hk according to the atom defect and nitrogen positions in the SiC crystals. For a 4H-SiC sample with high concentration of NV centers, the photoluminescence spectrum is shown in Figure 1(b). With off-resonant excitation (980nm), it is inevitable to excite all four NV centers within an excitation volume. Indeed, as shown in Figure 1(b), the PL spectrum shows the simultaneous presence and excitation of all four NV centers within the laser illumination volume. No evident of VV vacancy is excited under this condition for this sample and all other samples studied in this letter.

The zero-phonon line (ZPL) absorption spectra associated to both axial and basal color centers at 10K are further investigated under resonant excitation. For this experiment, we employ a fiber coupled, high resolution and tunable diode laser with laser power of 10μW. This low power level ensures that only one type of NV center is excited with negligible off-resonant excitation efficiency for the other NV centers, as seen from the background level in Figure 2(a). This excitation power is kept the same for all centers. The wavelength dependent count rate of the phonon side band (PSB) is recorded with a superconducting single photon detector with efficiency optimized at telecom wavelength. The excitation spectra of all four NV centers are shown in Figure 2(a). Interestingly, the two axial centers show a doublet structure in the excitation spectra with a small splitting of 0.7nm, whereas the basal ones show only a single peak. We label the doublet peaks for hh defects as PK1 and PK2, and those for kk defects as PK1' and PK2' respectively. According to [18,35], this doublet structure is related to the existence of two close excited states for axial NV centers. However, the exact origin of the doublet structure of the axial defects and its opportunities for spin manipulation still require further investigation.

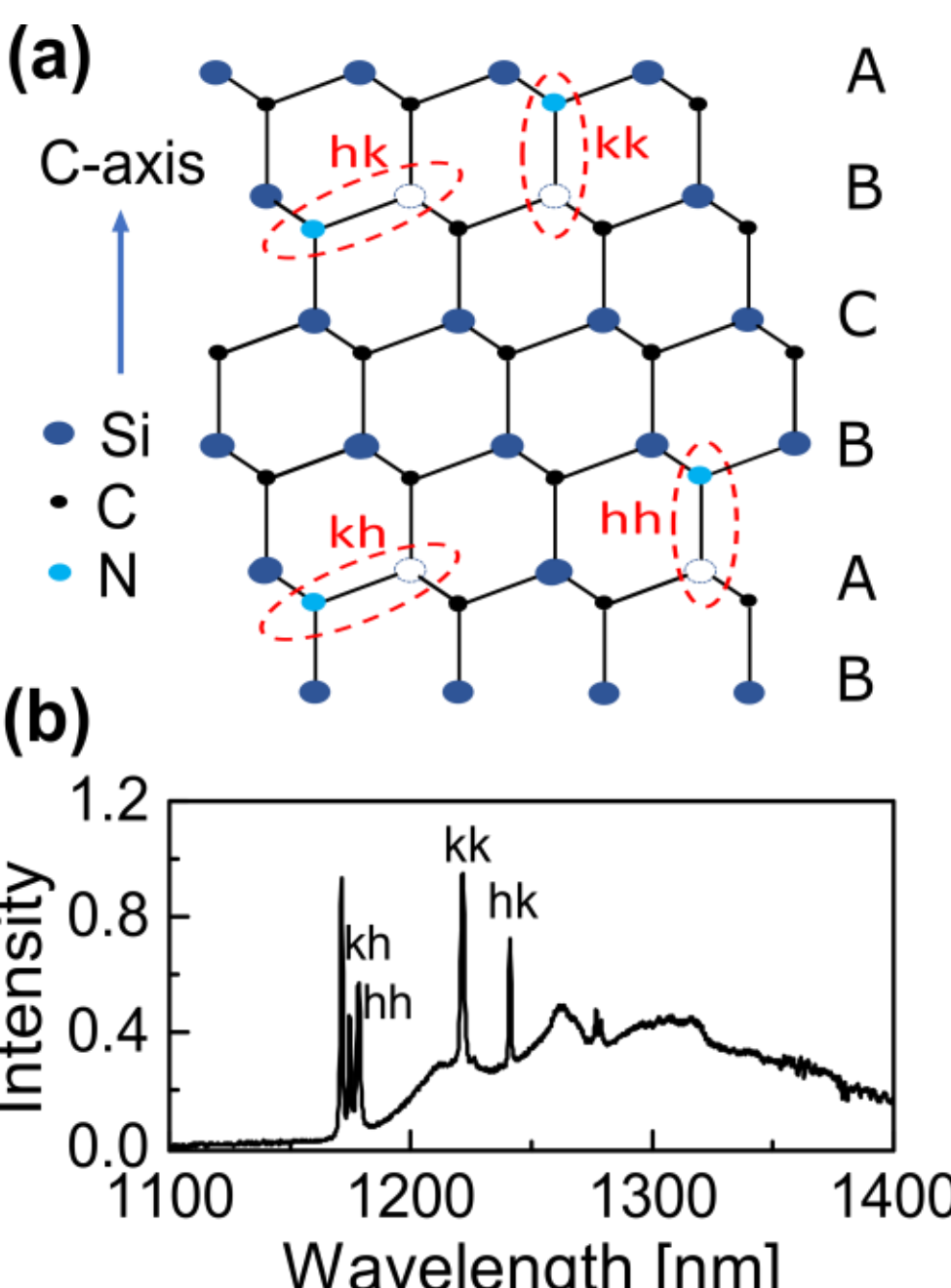

FIG. 1 NV center generation in 4H-SiC. (a) NV defects configurations in 4H-SiC. The four possible types of NV centers are indicated. The empty sites refer to Si vacancies and the light blue ones refer to the C substituted N

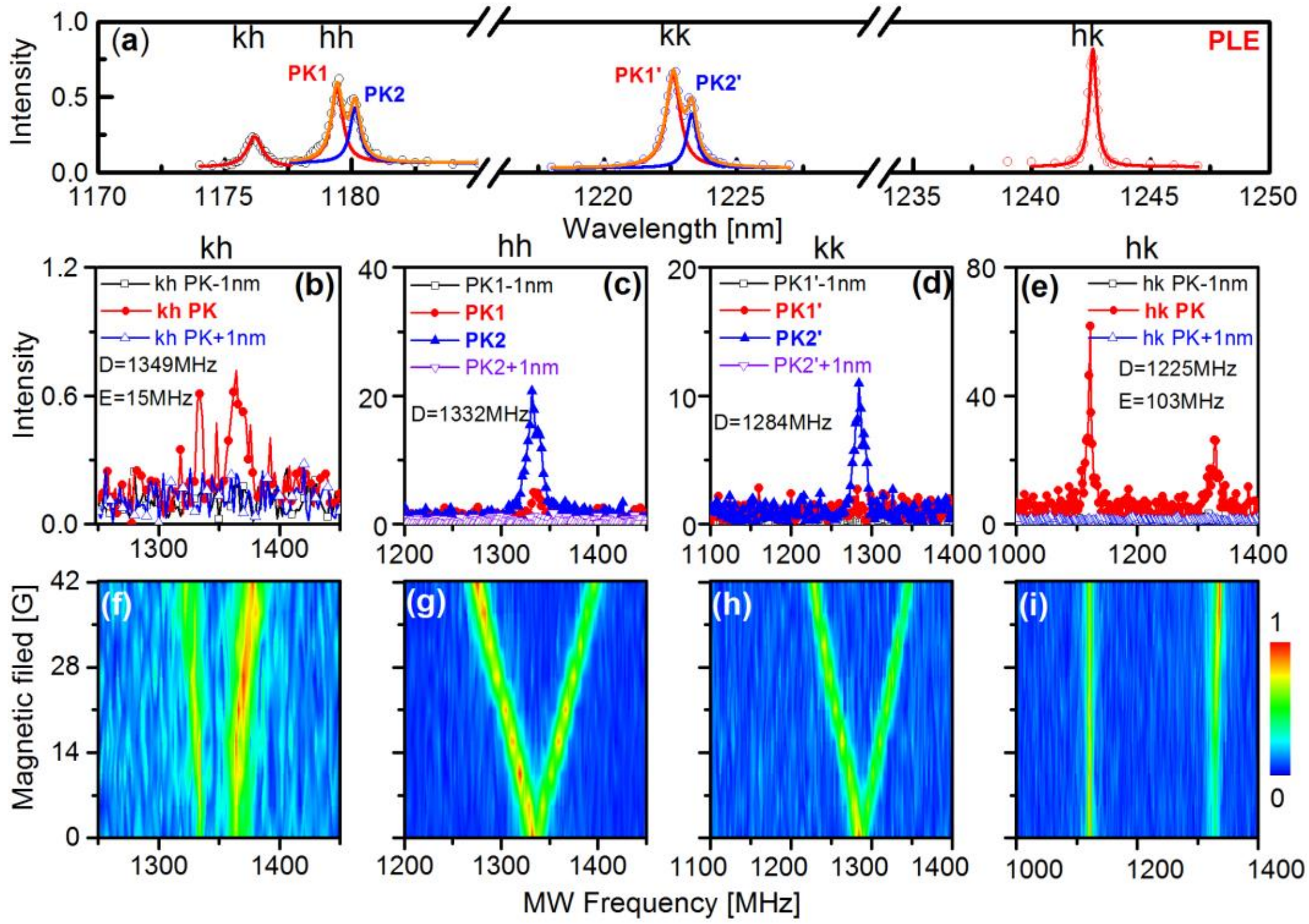

FIG. 2. Low temperature resonant excited PL and ODMR spectra. (a) Resonant excited PL of each NV center. The dots referred to the experimental results. The excitation spectra are fit with Lorentz functions. The peak absorption wavelength and linewidth are 1176.170nm/0.731nm (kh), 1242.571nm/0.366nm (hk), 1179.419nm/0.547nm (PK1 of hh) and 1180.130nm/0.497nm (PK2 of hh), 1222.596nm/0.641nm (PK1' of kk) and 1223.303nm/0.473nm (PK2' of kk). (b. c. d. e) Resonantly excited ODRM spectra at zero magnetic field. The ZFS parameters $D$ and $E$ of each NV center are shown in the figures. ODMR for on-resonant excitation and off-resonant excitation (with wavelength detuned by 1nm). ODMR spectra as a function of applied magnetic field parallel to the c-axis are shown in (f. g. h. i)..

neighbors. (b) Low temperature PL spectrum obtained for 980nm non resonant excitation at 10K. The four types of ZPL PL lines are indicated in the figure accordingly.

For the ODMR measurements we have deposited microwave ring antenna of typical size of 100μm on the sample surface to excite the Zeeman transitions of each center. For each ODMR experiment, we excite the NV centers resonantly while collect the PSB emission only. The laser excitation was filtered out with the help of a combination of 1250nm long-pass filter and a band-pass filter centered at 1300nm (bandwidth 50nm). For spin manipulation, the microwave is amplitude modulated by a function generator with a frequency of 134Hz and scanned across a certain frequency range. The time-tagged detector count rate is registered with Fourier transformation applied to the count rate to obtain the ODMR signal at 134Hz [8,36].

In Figure 2b-e the resonantly excited ODMR spectra are shown for all four NV centers. The microwave is swept from 1000 MHz to 1400 MHz and the power is kept at same level for each kind of center when laser is either on resonant with the center or is detuned from resonance. In this range we observe 6 resonances which we can associate with the different NV centers. As expected, the ODMR signal is strongest when the laser is resonant with the absorption spectra. The absorption spectra are very narrow with a width below 1nm. This high selectivity allows the association of the ODMR spectra with the different types of NV centers. The ground state spin properties for a state with spin S=1 can be described by the following Hamiltonian [10,37]:

$$H = \mu_B g \boldsymbol{B} \cdot \boldsymbol{S} + \hbar D\left(S_Z^2 - \frac{1}{3}S(S+1)\right) + \hbar E\left(S_X^2 - S_Y^2\right) \tag{1}$$

Where $\mu_B$ is the Bohr magneton, **g** is the Lande g-factor, the external magnetic field $\boldsymbol{B}$ coupled to the spin $\boldsymbol{S}$, and $D$ and $E$ are related to the axially symmetric and anisotropic components of the zero-field splitting (ZFS). For axial NV centers, the transverse anisotropy spin splitting ($E$) is zero and the ground state splitting is therefore determined by the axial term $D$ only. For basal NV centers, the ground state spins have a non-zero $E$ and result in two zero-field resonances at frequencies defined by ($D \pm E$). D and E values are obtained from the ODMR spectrum at zero field and indicated in each panel. The values of $D$ and $E$ and the comparison with the previously reported one's obtained by EPR measurements are summarized in Table SI in the supplementary information. Due to the higher resolution of ODMR spectroscopy these results lead to a refinement of the $D$, $E$ values.

We further studied the effect of an applied magnetic field on the ODMR resonances. In our experiment, an out of plane magnetic field up to 42 Gauss is applied, which is 4 degree deviated from the c-axis of the sample. For axial NV centers (hh and kk), a strong Zeeman splitting can be observed (Figure. 2g and 2h), which agrees with the known, microscopic models of the NV centers. For the basal NV centers, the ODMR spectrum shifts only weakly, which corresponds to the splitting $2\sqrt{\left(g\mu_B B/\hbar\right)^2 + E^2}$ , where $B$ represents the magnetic field projected onto the quantization axis of basal centers.

The doublet structure of the excitation spectra of the axial NV centers (hh and kk) has been observed in resonant excited EPR as well. However, whereas both peaks lead to a comparable ground state spin polarization in EPR spectroscopy, they show a different spin initialization behavior in ODMR (Figure 2c&d). When the laser wavelength is on resonance with the first peak (labeled as PK 1), the ODMR peak intensity (Figure 2c red line) is only 25.4±1.4% of the one when the laser is on resonance with the peak PK2 (Figure 2c blue line). For the kk NV center, the selectivity is even greater as the ODMR signal is only revealed when the laser wavelength is set to on resonance with PK2'. In other words, in ODMR the spin polarization is only observed when the lower of the two excited states is excited. Since the spin initialization is believed to be generated by the recombination via an intermediate singlet state, similar to the case of the NV center in diamond [6], our results may suggest a different coupling efficiency of the two excited states [6,18] to the metastable state and therefore different spin initialization efficiency.

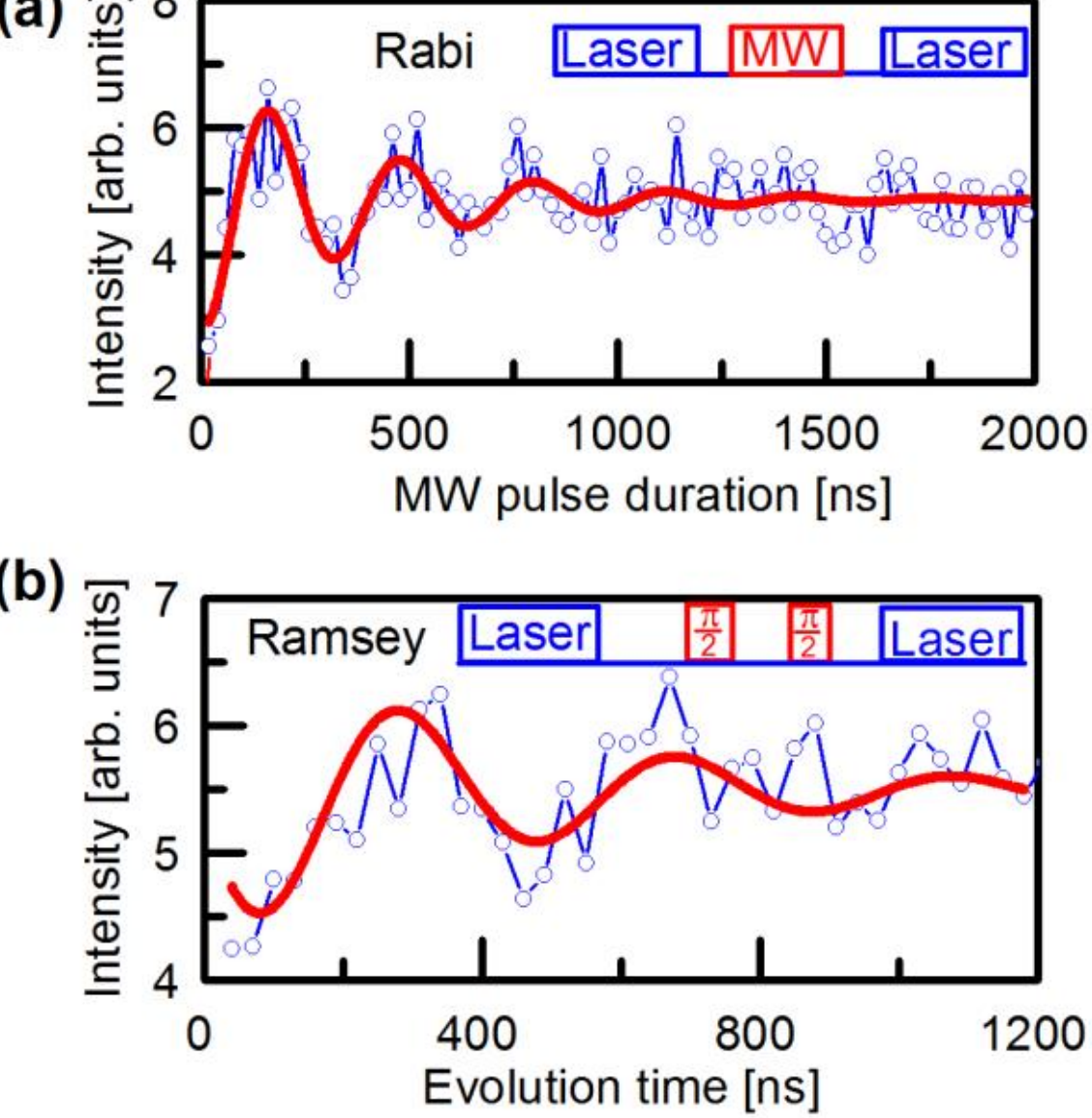

FIG. 3 Coherent manipulation of kk type NV color centers at 10K with magnetic field of 18 Gauss. (a) The spin Rabi oscillation. (b) The Ramsey fringe of electron spins in kk color centers.

A major parameter for the application of NV centers in quantum technology is their spin coherence, which can be investigated with Rabi and Ramsey experiments. In Fig. 3 we show the results for the axial kk NV center, which has the strongest PL spectrum. For these measurements we applied a weak magnetic field of 18 Gauss parallel to the c-axis to break the degeneracy of the $^3A_2$ $m_s$=1 spin state. The following measurements were performed on the higher energy $m_s$=1 state. The initialization and readout laser pulses are set to 5μs and the microwave pulse length is varied from 10ns to 2μs. Rabi oscillation with several oscillation periods are observed (Figure 3a), indicating the coherent manipulation of the kk NV centers. The experimental result is fit with a decaying sinusoidal function (equation 2), giving a Rabi frequency of 3.12±0.11MHz and a Rabi decay time of 376±24ns.

$$I = A + B\cos(2\pi f t)\exp(-t/T) \quad (2)$$

In order to estimate the dephasing time ($T_2$*) of the kk NV center, a Ramsey experiment with a 2MHz detuned microwave frequency is performed (Figure 3b). The $T_2$* related to the kk NV center is obtained by fitting the curve with equation (3). A fit to the experimental results gives a spin coherence time of $T_2$*= 388±136 ns.

$$I = A' + B'\cos(2\pi f' t + \varphi')\exp[-(t/T_2^*)] \quad (3)$$

This value is comparable to the case of divacancies in 4H-SiC, where for the kk VV in 4H-SiC a value of $T_2$* ~ 185ns has been reported [8]. These measurements were done on a proton irradiated, high concentration sample. The spin-spin interaction between the centers, is expected to shorten the dephasing time [22]. Much longer $T_2$* can be expected for single isolated defects.

Whereas proton irradiation is a convenient technique for the formation of homogeneous high concentration of NV centers, low energy (keV) ion implantation with nitrogen is a more appropriate for introducing locally and quantitatively controlled NV centers. We have applied this approach here. With large-area nitrogen implantation with 30*keV* and annealing at 1000°C for 1 hour, only NV center photoluminescence can be observed, as shown in Supplementary Figure 1. After carefully tuning the implantation dose to $10^{10}$/cm$^2$, single NV centers have been successfully identified, where we use laser excitation with 1039nm. As shown in Figure 4(a), the PL mapping of several emitters in a 100μm$^2$ area are shown with a long pass filter above 1200 nm. The single emitters were characterized using a Hanbury Brown and Twiss (HBT) interferometer. The normalized photon correlation raw data is shown in Figure 4(b) (blue line). We get $g^2(0) = 0.36 \pm 0.06 <$ 0.5, confirming the single photon emission properties. The non-ideal $g^2(0)$ value mainly comes from the affection of background counts. After background correction, we get a $g^2(0)$ value of $0.01 \pm 0.02$. Its saturation behavior and emission intensity stability curves are shown in Figure S3, which shows that they are stable single photon emitters. Additional single photon emission data can be found in Figure. S4.

The current single photon emission count rate is limited by the total reflection of surface and collection efficiency. In future research, it is desired to couple the created emitters into different types of optical cavities to increase their collection efficiency and study the light-matter interaction in strongly coupled regime. In this regard, the fabrication of emitters in a determined location is preferred. We demonstrate that it is possible to create arrays of NV centers in SiC in pre-defined positions by using focused ion beam direct writing. The confocal PL mapping and their PL spectrum are shown in Fig. S5, indicating the creations of NV centers in SiC with this method.

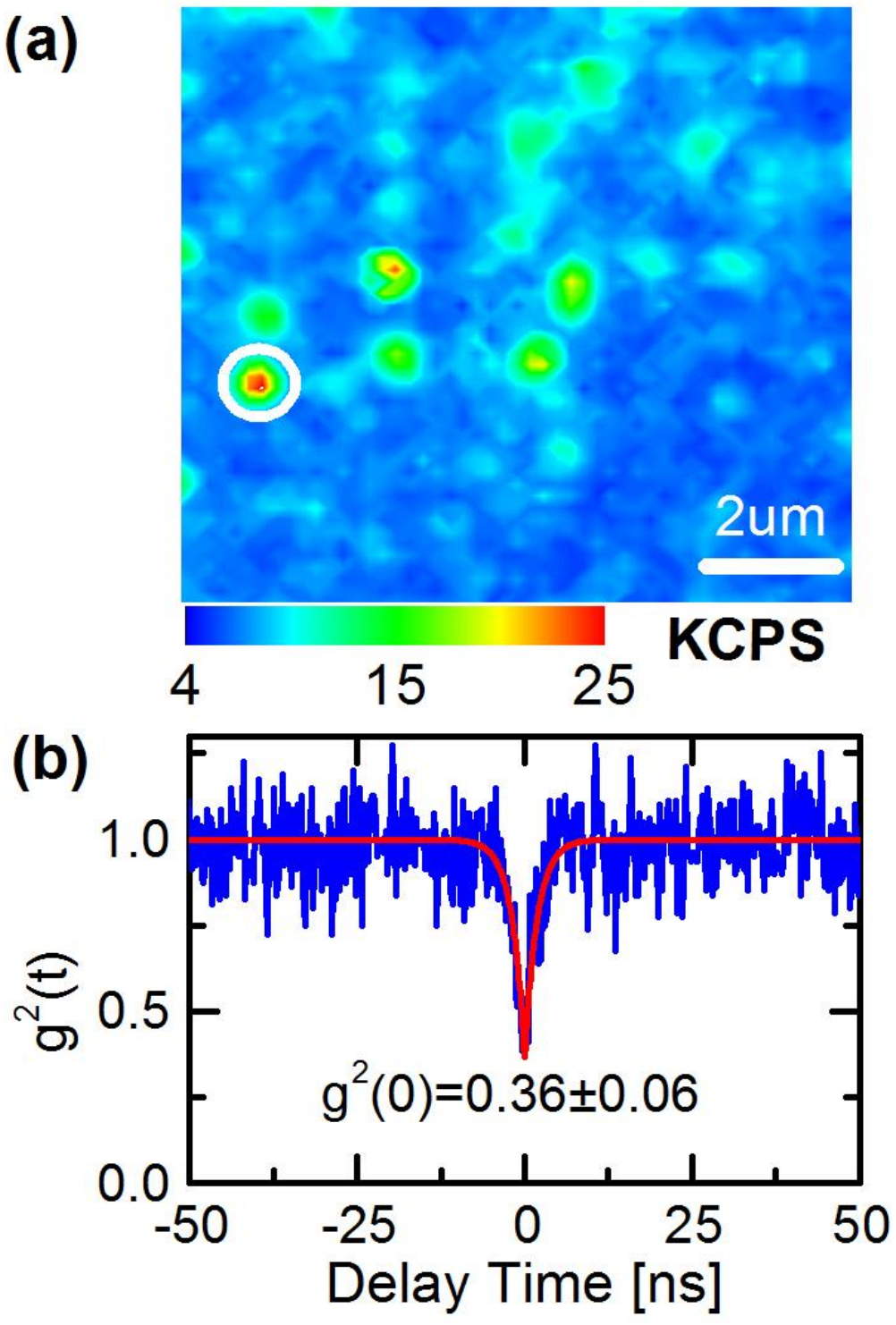

FIG. 4 Singe NV center generation. (a) Confocal mapping of a low dose ($10^{10}$/cm$^2$) implanted sample. The circled spot corresponds to the single emitter; the scale bar is 2μm. (b) Second order correlation measurement results without background correction (blue line). The red line is the fit, resulting a dip of 0.36±0.06.

In conclusion, we successfully performed ODMR spectroscopy of the four NV centers in 4H-SiC with resonant excitation, providing a method to tag and identify each of them. Taking advantage of this technique, we realized the coherent manipulation of electron spins of the NV center in 4H-SiC. We generated single NV centers with nitrogen ion implantation. With the still relatively weak emission rate of NV centers in 4H-SiC as compared to diamond, the incorporation of NV centers into nanostructures like solid immersion lens [13] or photonic crystal cavity [38] is desired for quantum technology applications. NV creation with the focused ion beam (FIB) implantation shows the promise to integrate NV into nanostructures for future quantum computation and quantum sensing researches.

**Acknowledgments**

We acknowledge the financial support from the Singapore National Research Foundation through a Singapore 2015 NRF fellowship grant (NRF-NRFF2015-03), Singapore Ministry of Education (MOE2016-T2-2-077, MOE2016-T2-1-163, MOE2016-T3-1-006 (S)), and A*Star QTE programme and Australian Research Council (DP180100070).

*Note added.* —Recently, Junfeng et al. reported a similar result to ours [39]. The off-resonant excitation is used there. Both works are complimentary and independent.

[1] L. Childress and R. Hanson, MRS Bulletin **38**, 134 (2013).
[2] M. Atatüre, D. Englund, N. Vamivakas, S.-Y. Lee, and J. Wrachtrup, Nature Reviews Materials **3**, 38 (2018).
[3] D. D. Awschalom, R. Hanson, J. Wrachtrup, and B. B. Zhou, Nature Photonics **12**, 516 (2018).
[4] C. Bradac, W. Gao, J. Forneris, M. E. Trusheim, and I. Aharonovich, Nature Communications **10**, 5625 (2019).
[5] C. L. Degen, F. Reinhard, and P. Cappellaro, Reviews of Modern Physics **89**, 035002 (2017).
[6] R. Schirhagl, K. Chang, M. Loretz, and C. L. Degen, Annu Rev Phys Chem **65**, 83 (2014).
[7] P. C. Maurer *et al.*, Science **336**, 1283 (2012).
[8] W. F. Koehl, B. B. Buckley, F. J. Heremans, G. Calusine, and D. D. Awschalom, Nature **479**, 84 (2011).
[9] A. Lohrmann, B. C. Johnson, J. C. McCallum, and S. Castelletto, Reports on Progress in Physics **80**, 034502 (2017).
[10] A. L. Falk, B. B. Buckley, G. Calusine, W. F. Koehl, V. V. Dobrovitski, A. Politi, C. A. Zorman, P. X. Feng, and D. D. Awschalom, Nat Commun **4**, 1819 (2013).
[11] D. J. Christle, A. L. Falk, P. Andrich, P. V. Klimov, J. U. Hassan, N. T. Son, E. Janzen, T. Ohshima, and D. D. Awschalom, Nat Mater **14**, 160 (2015).
[12] H. J. von Bardeleben, J. L. Cantin, E. Rauls, and U. Gerstmann, Physical Review B **92**, 064104 (2015).
[13] M. Widmann *et al.*, Nat Mater **14**, 164 (2015).
[14] H. J. von Bardeleben, J. L. Cantin, A. Csóré, A. Gali, E. Rauls, and U. Gerstmann, Physical Review B **94**, 121202 (2016).
[15] S. A. Zargaleh *et al.*, Physical Review B **94**, 060102 (2016).
[16] J. Wang *et al.*, ACS Photonics **4**, 1054 (2017).
[17] S. A. Zargaleh, S. Hameau, B. Eble, F. Margaillan, H. J. von Bardeleben, J. L. Cantin, and W. Gao, Physical Review B **98**, 165203 (2018).
[18] S. A. Zargaleh, H. J. von Bardeleben, J. L. Cantin, U. Gerstmann, S. Hameau, B. Eblé, and W. Gao, Physical Review B **98**, 214113 (2018).
[19] H. B. Banks, Ö. O. Soykal, R. L. Myers-Ward, D. K. Gaskill, T. L. Reinecke, and S. G. Carter, Physical Review Applied **11**, 024013 (2019).
[20] K. Khazen, H. J. von Bardeleben, S. A. Zargaleh, J. L. Cantin, M. Zhao, W. Gao, T. Biktagirov, and U. Gerstmann, Physical Review B **100**, 205202 (2019).
[21] R. Nagy *et al.*, Nat Commun **10**, 1954 (2019).
[22] W. F. Koehl, B. Diler, S. J. Whiteley, A. Bourassa, N. T. Son, E. Janzén, and D. D. Awschalom, Physical Review B **95**, 035207 (2017).
[23] L. Spindlberger *et al.*, Physical Review Applied **12**, 014015 (2019).
[24] H. J. von Bardeleben, S. A. Zargaleh, J. L. Cantin, W. B. Gao, T. Biktagirov, and U. Gerstmann, Physical Review Materials **3**, 124605 (2019).
[25] W. F. Koehl, B. Diler, S. J. Whiteley, A. Bourassa, N. T. Son, E. Janzén, and D. D. Awschalom, Physical Review B **95**, 035207 (2017).
[26] B. Diler, S. J. Whiteley, C. P. Anderson, G. Wolfowicz, M. E. Wesson, E. S. Bielejec, F. J. Heremans, and D. Awschalom, arXiv preprint arXiv:1909.08778 (2019).
[27] C. M. Gilardoni, T. Bosma, D. van Hien, F. Hendriks, B. Magnusson, A. Ellison, I. G. Ivanov, N. Son, and C. H. van der Wal, arXiv preprint arXiv:1912.04612 (2019).
[28] M. Widmann *et al.*, Nature materials **14**, 164 (2015).
[29] D. Simin, H. Kraus, A. Sperlich, T. Ohshima, G. Astakhov, and V. Dyakonov, Physical Review B **95**, 161201 (2017).
[30] R. Nagy *et al.*, Nature Communications **10**, 1954 (2019).
[31] V. A. Soltamov *et al.*, Nature Communications **10**, 1678 (2019).
[32] C. P. Anderson *et al.*, Science **366**, 1225 (2019).
[33] K. C. Miao *et al.*, Science Advances **5**, eaay0527 (2019).
[34] D. J. Christle *et al.*, Physical Review X **7**, 021046 (2017).
[35] A. Csóré, H. J. von Bardeleben, J. L. Cantin, and A. Gali, Physical Review B **96**, 085204 (2017).

[36] Y. Zhou, J. Wang, X. Zhang, K. Li, J. Cai, and W. Gao, Physical Review Applied **8**, 044015 (2017).
[37] L. Rondin, J. P. Tetienne, T. Hingant, J. F. Roch, P. Maletinsky, and V. Jacques, Reports on Progress in Physics **77**, 056503 (2014).
[38] T. Schroder *et al.*, Nat Commun **8**, 15376 (2017).
[39] J.-F. Wang *et al.*, arXiv preprint arXiv:1909.12481 (2019).

# Supplementary Materials

## Section I: Experimental methods

A home-built confocal microscopy is used for all characterization. Depending on different measurements purpose, different combinations of dichroic mirror or beam splitter (50:50), oil or air objective, long-pass or bandpass filter are employed. In detail, the room temperature $g^2$ measurement is performed with oil objective (N.A.=1.35, 100X). For second correlation measurement, the excitation laser (1039nm) is cleaned with 1050m short pass and the light collection is filtered with a combination of 1150nm dichroic mirror and 1200nm long-pass, and then it is guided to a superconducting single photon detector (Scontel). To acquire the emission spectra related to these samples (NV ensembles mounted in Montana Cryostation), IR objective (Olympus, N.A=0.7, 50X) is used for photon collection. For spectrum measurement, the excitation (980nm) is cleaned with 1000nm short-pass while a long-pass of 1000nm is applied in collection arm so as to collect both the possible emission related to divacancy and NV centers (InGaAs detector operating till 1.7μm is used). In order to realize resonant excitation of each type of NVs, a 50:50 beam-splitter (1100-1600nm) is applied. A monochromatic diode laser (Sacher) is used for on-resonant excitation and the laser is cleaned with a proper bandpass filter. For all CW and pulsed ODMR characterization, a combination of 1250nm long-pass and a bandpass of 1300nm (bandwidth of 50nm) is used for phonon side band collection. The collected photon is further guided to the superconducting single photon detector for time domain photon count.

The microwave feed through is realized with either a stripe or a ring fabricated on the sample whose gold pads are wire bonded to a chip carrier. For continuous ODMR test, a microwave source (Rohde&Schwarz SMIQ) is used. The microwave is then gated by microwave switch (ZASWA-2-5-DRA+) controlled by a function generator (modulation frequency of 134Hz). Before feeding into the Cryostation, the microwave is amplified with a microwave amplifier (TVA-4W-422A+). The time tagged count acquired by SSPD is then Fourier transformed so as to get the signal at each microwave frequency. For Rabi and Ramsey measurement, a second switch after the first one is applied. The second switch is controlled by a pulse plaster to produce pulses for spin control [2,3].

## Section II: Sample preparation

In order to obtain a high concentration of NV centers and a homogenous defect distribution over the entire thickness of the sample, we used N doped ($2x10^{17}/cm^3$) bulk samples which were irradiated with 12MeV protons at a fluency of $10^{16}/cm^2$. After high temperature annealing (900°C), the sample contained NV centers in the $10^{16}/cm^3$ concentration range [4]. For ODMR measurements, microwave antenna were deposited on the surface of the sample in a conventional manner [2].

Those substrates which are implanted with Nitrogen ion beam, either with ion implantation (30keV) or with focused ion beam (30keV), are commercially available substrate purchased from Norstel (high purity bulk 4H-SiC epitaxies on the silicon face). These samples are low dose implanted so as to realize single NV search. The dose per site is controlled by the ion fluency and the dwell time. The prepared sample is then loaded into a high vacuum environment ($2x10^{-6}$ mbar) for annealing 1h so as to activate the NV centers [5,6]. The annealing temperature are 800 °C and 1000 °C. After annealing, the samples are clean with hot Piranha acid (150 °C) for 1 hour.

| | EPR [MHz] | | ODMR [MHz] | |
|---|---|---|---|---|
| | D | E | D | E |
| kk | 1282 | 0 | 1284 | 0 |
| hh | 1331 | 0 | 1332 | 0 |
| hk | 1193 | 104 | 1225 | 103 |
| kh | 1328 | 15 | 1349 | 15 |

Table SI: D and E values comparison with resonant EPR and ODMR characterization. Overall, the results are in agreement with each other except that the D values acquired with ODMR method for basal center is greater compared to the EPR ones [1]. Different temperature needs to be considered for the difference.

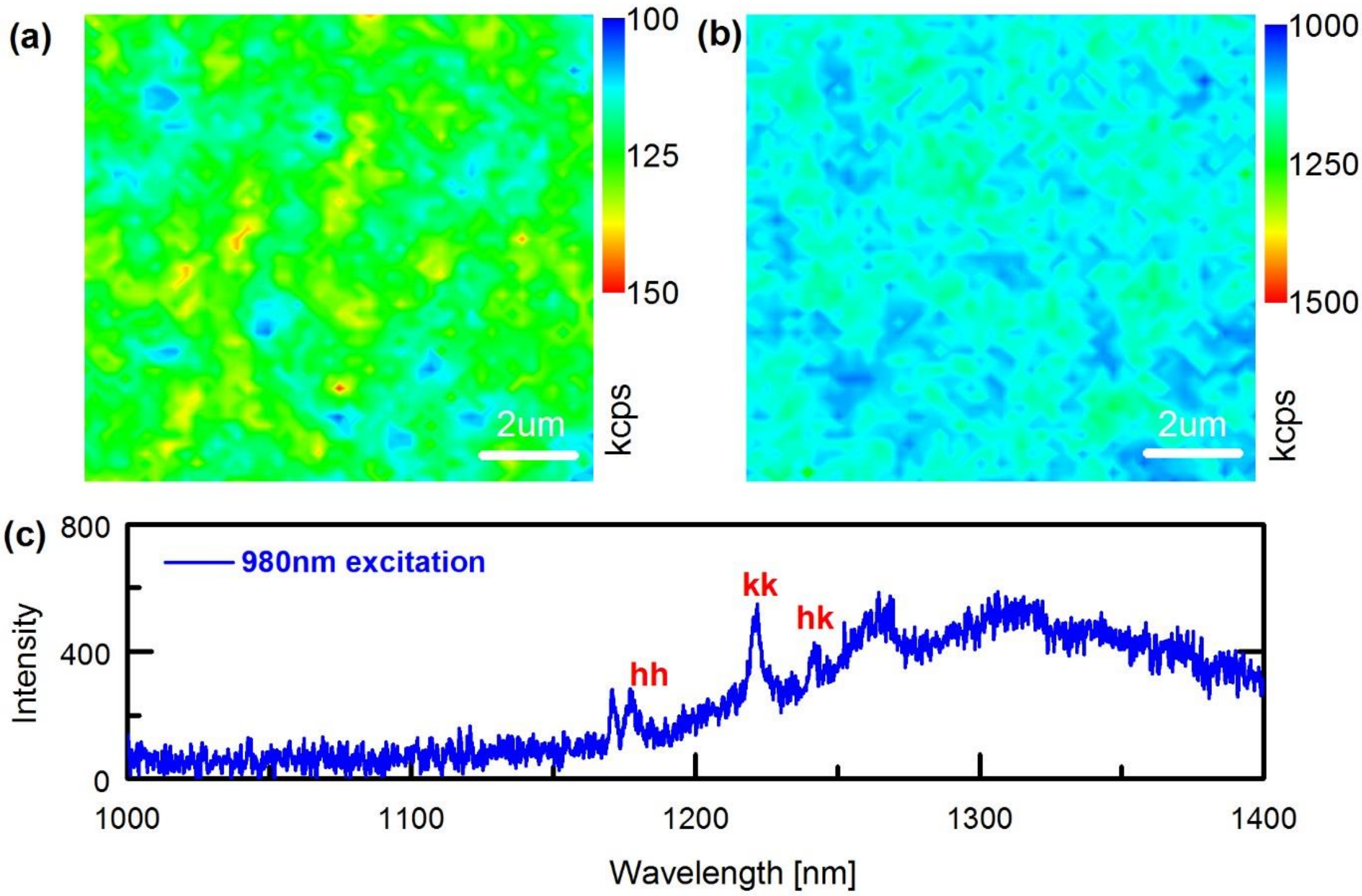

FIG. S1. 2D confocal PL mapping of nitrogen implanted 4H-SiC samples. (a) 2D mapping for sample with implantation dose of $10^{13}/cm^2$. The maximum count rate is 150K/s. (b) 2D mapping for sample with implantation dose of $10^{14}/cm^2$. The maximum count rate is 1500K/s. (c) Spectrum of the sample in (b) under 980nm excitation at 10K. The low temperature spectrum of the Nitrogen irradiated sample clearly proves the creation of NV centers. Meanwhile, there is no evident peaks related to divcancy. According to the 2D

mapping, the count rate is proportional to the implantation dose. This then makes single NV generation at low dose implantation possible.

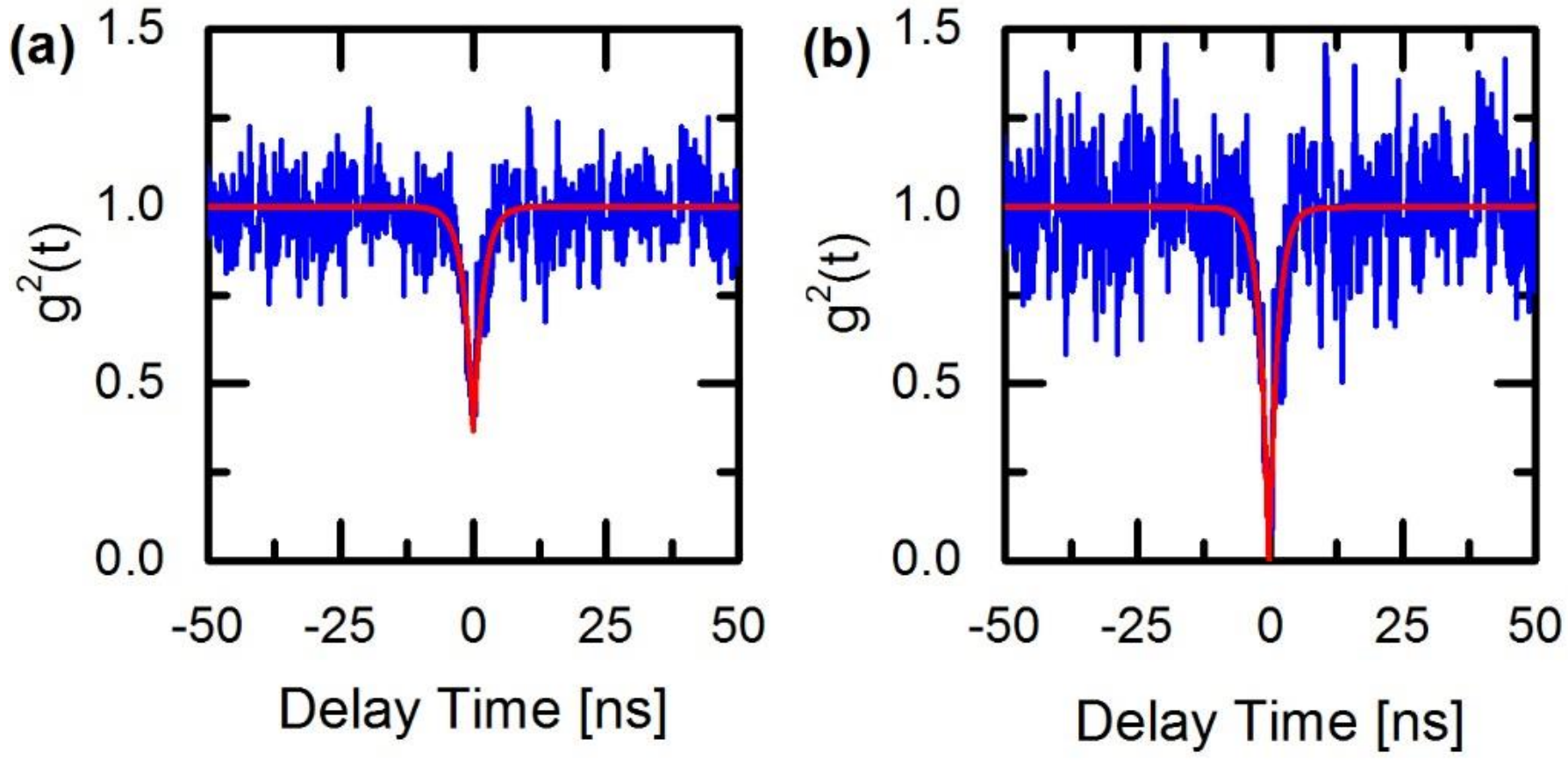

FIG. S2. $g^2(0)$ value raw data verse background corrected data. (a) $g^2$ raw data (blue line) and its fit (red line) with $g^2(t) = A + B\exp(-|t|/\tau_0)$ giving a dip of 0.36±0.06 and $\tau_0$=1.74±0.22ns. (b) $g^2$ data with background corrected (blue line) and its fit (red line) giving a dip of 0.01±0.02. The raw data is corrected with $g^2(t) = [N(t)/N_n - (1-\rho^2)]/\rho^2$, where $\rho = S/(S+B)$, with S refers to signal and B background respectively [7]. Here, we have signal count rate of 16K/s and background level of 4K/s.

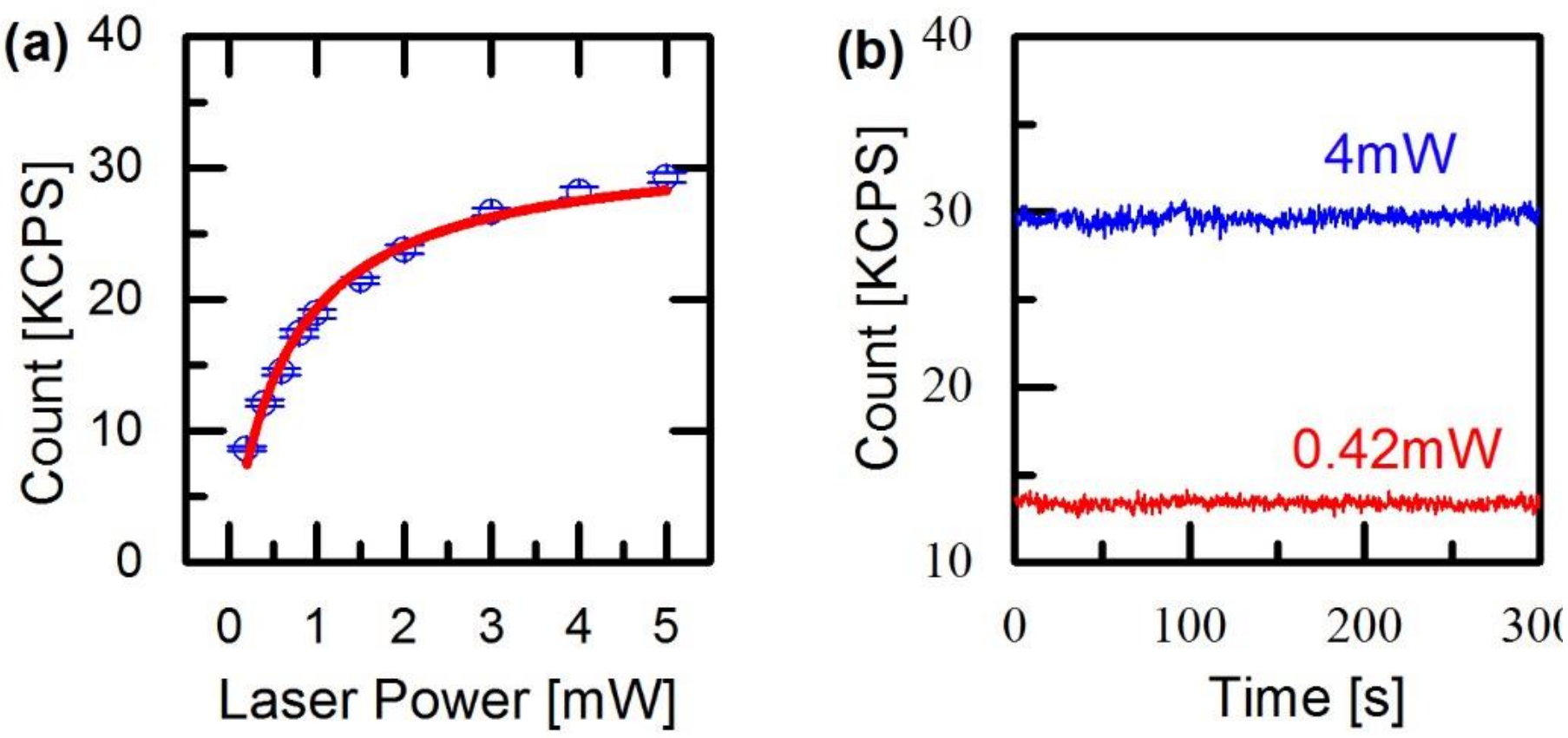

FIG. S3. The saturation behavior of the NV emitter. (a) Saturation curve of the learnt NV single emitter in the main text. The saturation count is 27.1±1.5K, and the saturation power is 657.4±1.6μW. (b) time trace of the emission with under saturated and saturated excitation, showing the stable emission of the NV center.

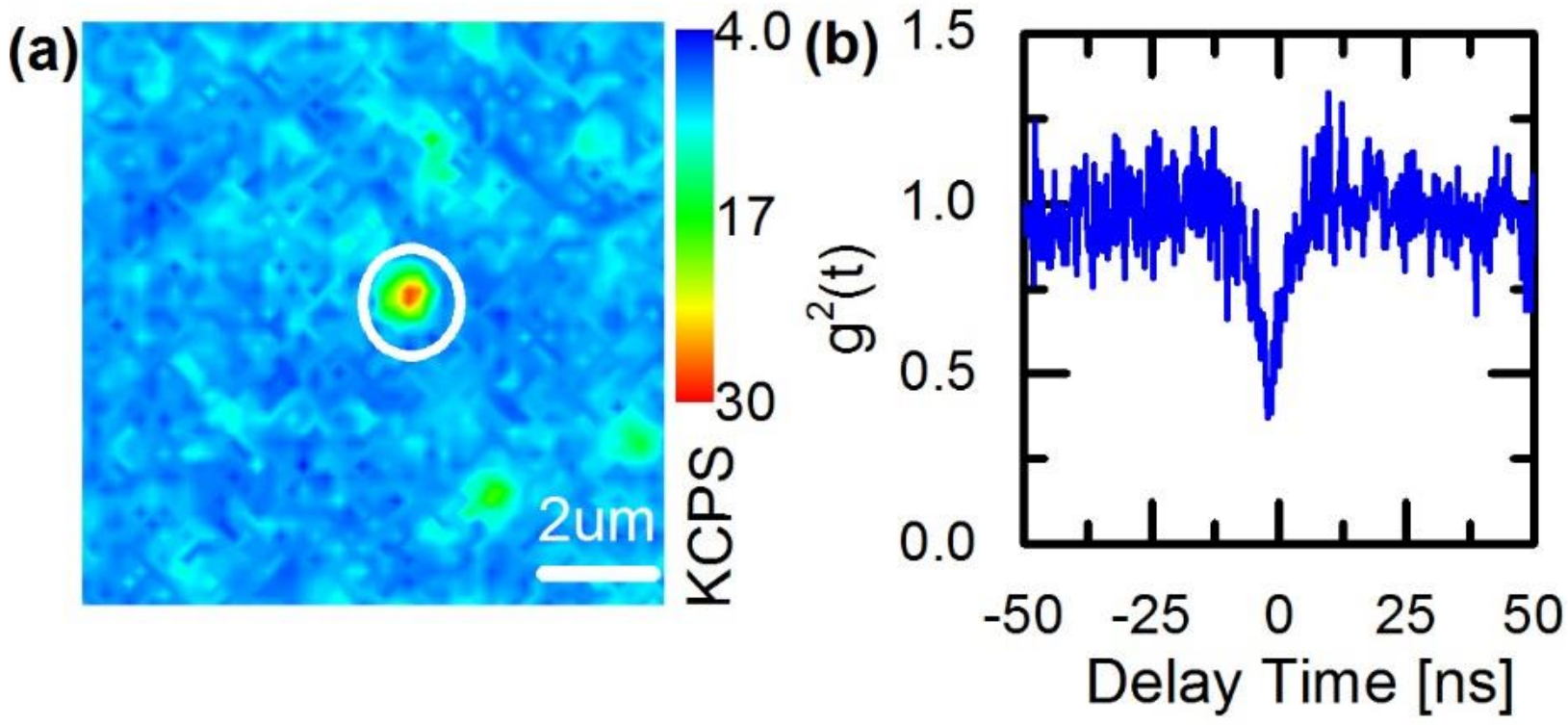

FIG. S4. Additional single NV emitter. (a) Confocal PL map of the same sample used in Main text. The scale bar is 2μm. The circled spot refers to the one with which the HBT is performed. (b) $g^2$ measurement of the circled spot. The blue curve is the raw data without background correction, according to which the dip is lower than 0.5 (0.39) and confirms the single emission of the circled spot. As a high power of 4mW is applied, there is a bunching behavior due to the shelving state in the excited states.

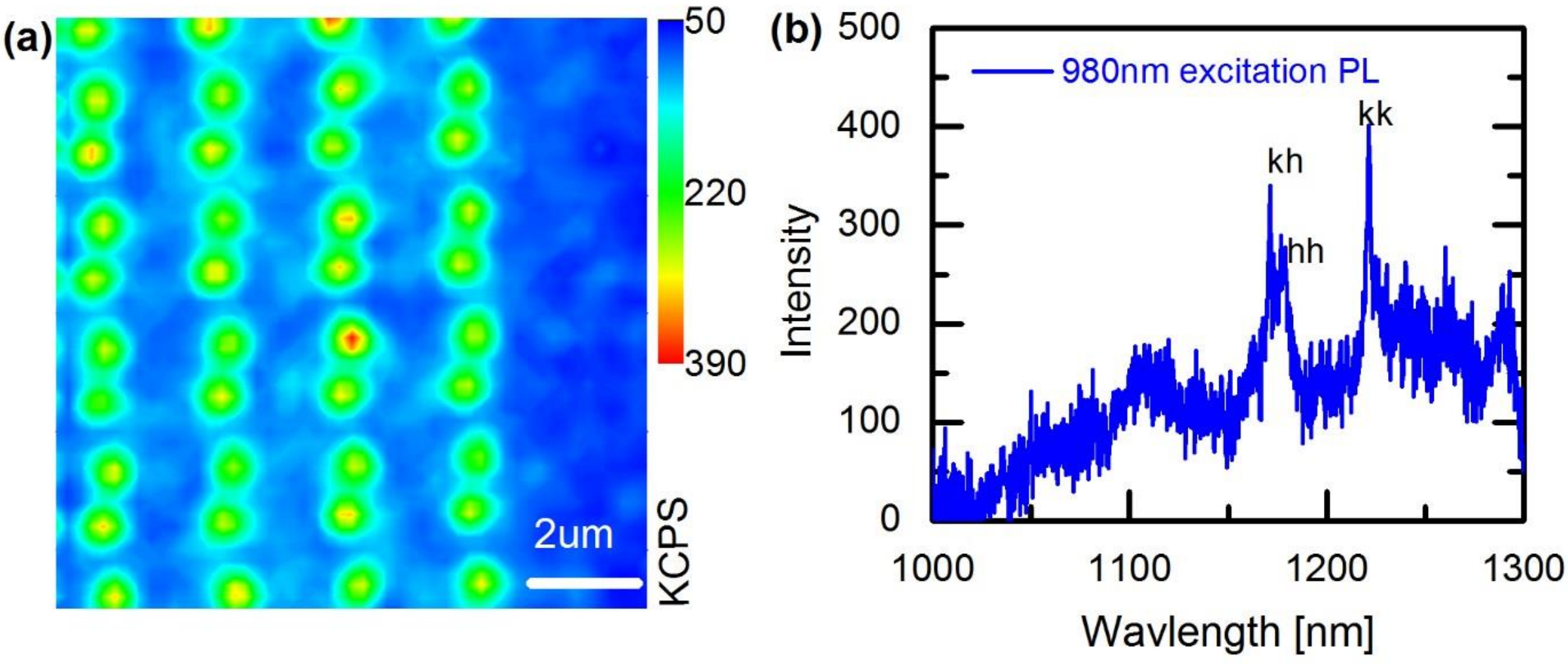

FIG. S5. NV center generation with focused ion beam writing. (a) Confocal mapping of focused nitrogen ion beam implanted sample at room temperature. The scale bar is 2μm. Due to the imperfection of Nitrogen beam, one shot resulted in two implanted spots. (b) Spectrum related to the focused ion implantation sample at 10K with 980nm laser excitation and collection above 1000nm. The peaks related to each kind of NV centers are labeled accordingly.

[1] S. A. Zargaleh, H. J. von Bardeleben, J. L. Cantin, U. Gerstmann, S. Hameau, B. Eblé, and W. Gao, Physical Review B **98**, 214113 (2018).

[2] Y. Zhou, J. Wang, X. Zhang, K. Li, J. Cai, and W. Gao, Physical Review Applied **8**, 044015 (2017).

[3] W. F. Koehl, B. B. Buckley, F. J. Heremans, G. Calusine, and D. D. Awschalom, Nature **479**, 84 (2011).

[4] H. J. von Bardeleben, J. L. Cantin, E. Rauls, and U. Gerstmann, Physical Review B **92**, 064104 (2015).

[5] Journal of Applied Physics **126**, 083105 (2019).

[6] S. A. Zargaleh *et al.*, Physical Review B **94**, 060102 (2016).

[7] D. J. Christle, A. L. Falk, P. Andrich, P. V. Klimov, J. U. Hassan, N. T. Son, E. Janzen, T. Ohshima, and D. D. Awschalom, Nat Mater **14**, 160 (2015).